\documentclass[a4paper,11pt]{article}
\usepackage{pos}

\usepackage{array}
\usepackage{subcaption}

\title{Recent Developments in Colibri: A Neural-Network PDF Model and Improved Bayesian Sampling}

\author[a]{Ella~Cole}
\author*[a]{Valentina~Sch\"utze~S\'{a}nchez}

\affiliation[a]{DAMTP, University of Cambridge, Wilberforce Road, Cambridge, CB3 0WA, UK}

\emailAdd{egc41@cam.ac.uk}
\emailAdd{vcs32@cam.ac.uk}

\abstract{\noindent We present recent developments in Colibri, an open-source framework for flexible fits of parton distribution functions (PDFs). Colibri allows the implementation of any PDF functional form as well as Bayesian, Monte Carlo, and Hessian inference methods, making it particularly well suited for systematic comparisons of model architectures and uncertainty propagation treatments. In these proceedings, we report on two key updates to the framework: the release of \texttt{colibri-n3fit}, an open-source implementation of the neural-network model used by the NNPDF collaboration, and the integration of an upgraded, highly efficient Bayesian sampler, bridging the gap between the computational cost of Bayesian PDF fitting and other uncertainty propagation treatments. Combined, these updates extend Colibri's capabilities to neural-network PDF fits, enabling direct, quantitative Bayesian analyses of the \texttt{n3fit} architecture and opening new avenues for comparative studies.}

\FullConference{43rd International Conference on High Energy Physics (ICHEP 2026)\\
30 July  to 5 August , 2026\\
Natal, Brazil\\}

\begin{document}
\maketitle

\section{Introduction}
\label{sec:intro}

\noindent Theory predictions for the proton structure are relevant not only in furthering our understanding of stable matter, but also as an important ingredient in new physics searches and in obtaining predictions for observables involving hadrons. In turn, the only way to date to constrain the proton’s structure is to leverage the high volume of precise data that is produced and collected at CERN’s Large Hadron Collider (LHC) and other Deep Inelastic Scattering (DIS) experiments, by applying the most accurate and revealing statistical treatment. Yet, tackling this statistical inverse problem allows for much freedom of parametrisation, inference methodology, and uncertainty treatment~\cite{Amoroso:2022eow,Del_Debbio_2022,Ubiali:2024pyg,Chiefa:2025loi}. 

Polynomial parametrisations of different forms have been widely adopted, for example by the MSHT~\cite{Bailey_2021,qcd-1999,hm-1984}, CT~\cite{Hou_2021} or JAM~\cite{Moffat:2021dji} collaborations, while the NNPDF~\cite{NNPDF:2021njg,NNPDF:2021uiq} collaboration uses a neural network, and other alternatives have been proposed, such as the use of a linear model~\cite{Costantini:2025wxp} or Gaussian processes~\cite{Candido:2024hjt}. Crucially, different parametrisation and methodological choices lead to different predictions in phenomenologicaly relevant observables. For example, recent calculations of the central value of the $W$ boson mass produced with different published PDF sets have been shown to differ by more than the nominal PDF uncertainty~\cite{2026wbosonmass}. This leads to the need to define a more statistically rigorous component of the PDF uncertainty due to the choice of parametrisation itself. A systematic comparison of PDF parametrisations is therefore essential, including equal dataset inputs and theory assumptions. 

Furthermore, the inference method to find parameters that best describe the data, and the uncertainty propagation treatments to apply are also a non-trivial choices. A combination of gradient descent with a Monte Carlo (MC) replica method or Hessian error propagations has been, so far, widely employed methodologies. But with studies showing that the Monte Carlo replica method produces posterior distributions that disagree with the Bayesian estimations in the presence of quadratic contributions~\cite{costantini2024, Costantini:2026sel}, a Bayesian approach has emerged as a desirable alternative to provide a robust uncertainty treatment, with the caveat that it is computationally expensive, to the point that, until recently, it was considered unfeasible.

While there are well established approaches to convert between Hessian and MC PDF sets~\cite{Carrazza:2015aoa, Cruz2026, Gao:2013bia,Carrazza:2016htc, Watt:2012tq}, and there have been various benchmarks and PDF combinations, such as the recent ones by PDF4LHC~\cite{Butterworth:2015oua,PDF4LHCWorkingGroup:2022cjn} or the MSHT and NNPDF combination of approximate N3LO sets~\cite{Cridge:2024icl}, in practice, benchmarks that keep up with the divergent fitting methodologies used by different collaborations are hard to achieve on a regular basis. Tools such as xFitter~\cite{Zenaiev:2016jnq, xFitter:2022zjb, xFitterDevelopersTeam:2017xal} have come a long way in providing a framework for systematic comparisons, but they are limited in the available parametrisations to compare, and focus on already used error propagation strategies, so are ill-suited to continue to address the ongoing developments within the PDF fitting community.

With these challenges in mind, Colibri has been presented as a tool that addresses the flexibility and complexity of the PDF fitting problem~\cite{2026colibri}. As a modular code that allows the implementation of any PDF parametrisation, systematic model comparisons are facilitated. Furthermore, access to the Bayesian model evidence allows for a quantitative comparison of parametrisations with the same dataset inputs. The MC replica method and Hessian error propagations are also available, readily allowing studies that isolate the methodology comparison aspect. All in all, it is a tool that addresses the complexity of the problem, and is capable of adapting to the new frontiers within PDF fitting.

These proceedings present unpublished, recent updates in the code, the motivation behind them, and the impact on the PDF fitting capabilities of Colibri. In Section~\ref{sec:n3fit} we share a benchmark of the now publicly available \texttt{colibri-n3fit} model in Colibri, and in Section~\ref{sec:blackjax} we present the implementation of a new Bayesian sampler, BlackJAX, and describe some of its properties. In Section~\ref{sec:conclusion-outlook} we discuss ongoing directions for further development within the code.

\section{Neural-Network Model}
\label{sec:n3fit}

\noindent The \texttt{colibri-n3fit} model inherits from the \texttt{n3fit} PDF parametrisation, developed by the NNPDF collaboration~\cite{NNPDF:2021njg,NNPDF:2021uiq}. This parametrisation uses a neural-network architecture to provide a flexible representation of the PDFs, with the aim of reducing the functional-form bias associated with more restrictive, traditional parametrisations.

The \texttt{n3fit} model defines a PDF for a given flavour, $k$, as a product of a preprocessing function and a neural network, as follows:

\begin{equation}
    xf_k(x, Q_0^2) = A_k x^{1-\alpha_k} (1-x)^{\beta_k} \mathrm{NN}_k(x;\bf{\theta})  
\end{equation}

\noindent Here, $A_k$ is a normalisation constant, added to ensure that the PDF satisfies sum rules, $\alpha_k$ and $\beta_k$ are preprocessing parameters which
are randomly sampled from a uniform distribution at the start of each replica fit, and $\mathrm{NN}_k(x;\bf{\theta})$ is a neural network with trainable parameters, $\bf{\theta}$.

When fitting the \texttt{colibri-n3fit} model using the Colibri framework, additional model-specific
options are available in the fit runcard. The \texttt{nodes} and
\texttt{activations} options specify the number of nodes and the activation
function in each network layer, respectively. Both have default values and may
be omitted, or set explicitly to customise the network architecture. The existing parameter initialisation setting is also extended with the
\texttt{glorot\_norm} option, which applies Glorot normal initialisation to the
network weights and sets the biases to zero. 

A series of closure tests have been performed to validate the implementation of this model within Colibri. 
These tests involve generating a full set of DIS pseudo-data at two levels: level-0, where each replica is generated from the same central pseudo-data values, with independent Monte Carlo fluctuations applied to each replica, and level-1, where the central pseudo-data values are first shifted by some noise which is drawn from the experimental covariance matrix before independent Monte Carlo fluctuations are applied to each replica\footnote{Producing pseudo-data using the \texttt{closure\_test\_level:1} and fit with the \texttt{run\_monte\_carlo\_fit} method is equivalent to producing level-2 data, as defined in NNPDF, where MC noise is also added onto the shifted central value.}. The \texttt{n3fit} model is then fitted to the resulting pseudo-data using both the NNPDF and Colibri frameworks, allowing their results to be compared. The DIS dataset contains $N_{dat}= 3089$ data points after kinematic cuts, and is described in detail in ~\cite{NNPDF:2021njg}.

\begin{figure}[t]
    \centering
    \begin{subfigure}{0.45\textwidth}
        \centering
        \includegraphics[width=\linewidth]{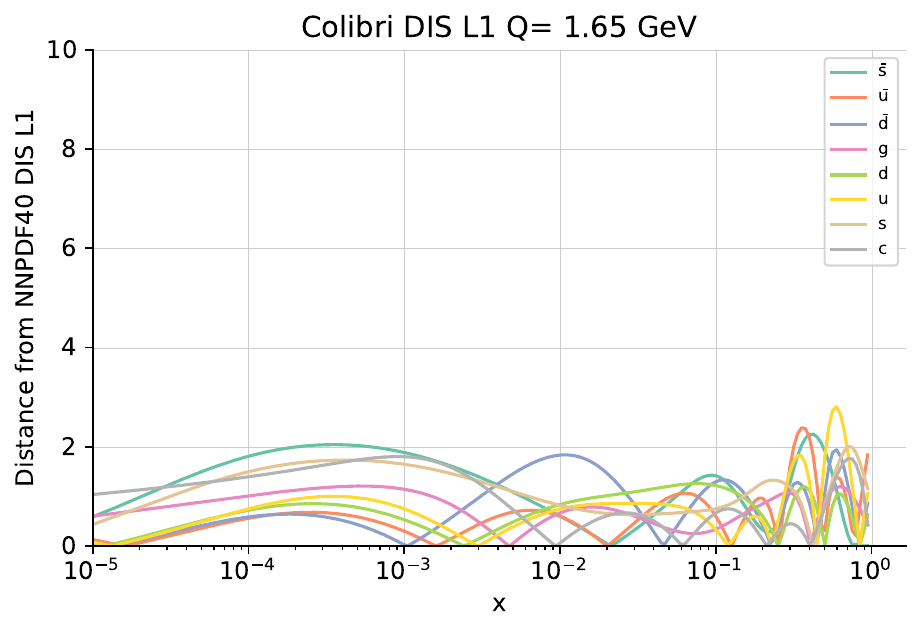}
        \caption{}
    \end{subfigure}
    \hfill
    \begin{subfigure}{0.45\textwidth}
        \centering
        \includegraphics[width=\linewidth]{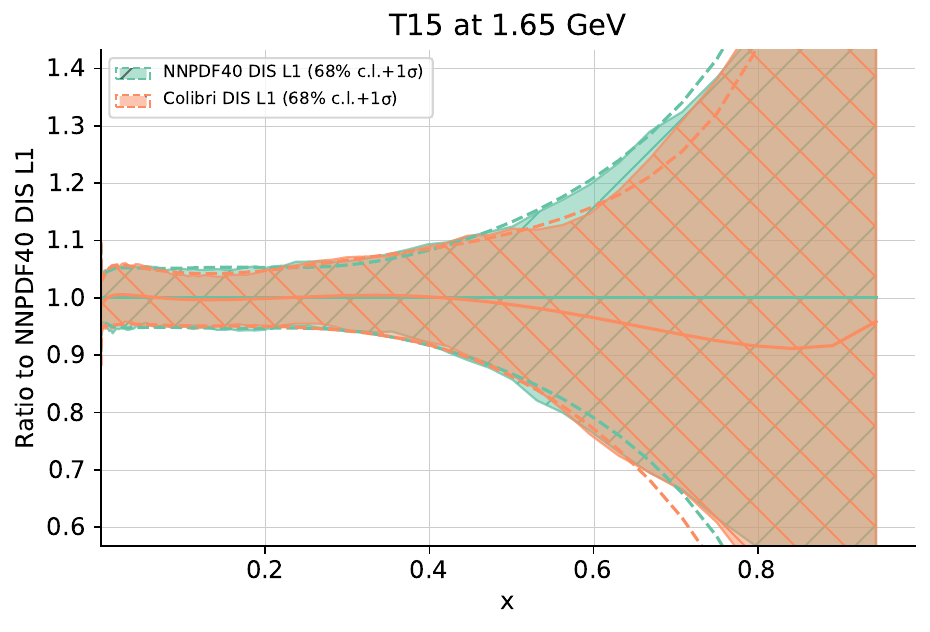}
        \caption{}
    \end{subfigure}

    \caption{Comparison of NNPDF and Colibri level-1 closure tests using \texttt{n3fit}: (a) distances between the central values of the evolution-basis PDFs, and (b) ratio of the non-singlet $T_{15}$ PDF fitted using NNPDF (blue) and Colibri (orange). Both plots are shown at the parametrisation scale, $Q = 1.65$ GeV.}
    \label{fig:n3fit}
\end{figure}

Figure~\ref{fig:n3fit} compares the level-1 closure-test results obtained with NNPDF and Colibri, showing the distances between the central values of the evolution-basis PDFs alongside the ratio of the non-singlet $T_{15}$ PDFs.
The distance is defined in \cite{NNPDF:2017mvq}, where $d=10$ corresponds to a $1\sigma$ deviation, while $d=1$ indicates statistical equivalence, consistent with replicas drawn from the same underlying probability distribution. The distances fluctuate around $d=1$, reaching maximum values of approximately $d=3$ in certain regions of $x$, indicating excellent agreement between the two frameworks. This agreement is further supported by the associated $\chi^2$ values, which are consistent across all fitted datasets. Similar agreement is also observed for the level-0 closure test results.

The implementation of the \texttt{n3fit} model in Colibri therefore provides a framework for performing Bayesian and Hessian analyses of PDFs parametrised using neural networks, enabling further studies of the impact of the fitting methodology on the resulting PDFs. Additionally, its implementation within a common framework allows for direct comparisons with alternative PDF models with the use of consistent data and theory settings.

\section{BlackJAX -- new nested sampler}
\label{sec:blackjax}

\noindent At the time of its original release, Colibri supported Bayesian nested sampling with UltraNest, which has the benefit of being a reliable, well-documented and maintained sampler~\cite{buchner2021ultranestrobustgeneral, Buchner_2014, Buchner_2019}. However, recent updates on alternative Bayesian sampling that are compatible with more parallelisation in computational operations have motivated the addition of BlackJAX as an alternative sampler~\cite{cabezas2024blackjax, yallup2025nested}.  

Compared to UltraNest, BlackJAX has the following advantages:
\begin{itemize}
    \item \textbf{GPU compatibility} UltraNest is GPU compatible, but runs some CPU $\leftrightarrow$ GPU operations in the background when running on GPU. BlackJAX can run entirely on GPU.
    \item \textbf{JAX optimisation} UltraNest can call JAX likelihoods, but uses \texttt{numpy} for all its internal calculations. BlackJAX runs entirely on JAX. 
\end{itemize}
This has the overall effect of making BlackJAX a much more computationally efficient Bayesian sampler. To illustrate this, we performed a comparison using the Les Houches parametrisation~\cite{Alekhin:2005xgg}, which is a polynomial model with 13 free parameters\footnote{This model is available to the user in the Colibri code as an example and as a template for other models.}. For this comparison, we performed a level-1 closure test, where pseudo-data has been generated by shifting the central values of some known, input theory prediction or \textit{underlying law}, by some noise which is drawn from the experimental covariance matrix, and this pseudo-data is the one used in the fit and compared to the underlying law.

We carried out comparisons with fits including the full set of deep inelastic scattering data used in NNPDF 4.0, which comprises 3089 data points. The fits were performed on a CPU with 4 allocated cores with a total memory of 16GB, and GPU with 2 allocated GPUs with 32 CPU cores each (which would be used by UltraNest but not BlackJAX) and a total memory of 500GB. In both cases, the samplers explored a vectorised likelihood up to 1\% precision, and with 500 live points per step (number of points drawn from the prior that are evaluated at a given step). The fits produced statistically equivalent results, with a $\chi^2$ of 0.998 in all four cases. This is compatible with a target $\chi^2$ of 1, corresponding to the normalised shift from the pseudo-data to the central values of the underlying law. Table~\ref{table:blackjax-vs-ultranest} shows a comparison between the running times for CPU and GPU. 
\begin{table}
\begin{center}
\renewcommand{\arraystretch}{1.2}
\begin{tabular}{|>{\centering\arraybackslash} m{7em} |c|c|>{\centering\arraybackslash}m{9em}|}
    \hline
     & UltraNest (UN) & BlackJAX (BJ)  & Relative runtime (UN/BJ) \\[0.5ex] 
     \hline
     \hline
     CPU runtime & 3784 s & 884.5 s & \textbf{4.28}\\\hline
     
     GPU runtime & 5275 s & 30.39 s & \textbf{174} \\\hline
     
     Relative runtime (CPU/GPU) & \textbf{0.72} & \textbf{29.1}& \\
     \hline
\end{tabular}
\caption{Comparison between the running speed of UltraNest and BlackJAX on CPUs and GPUs for level-1 closure tests with the Les Houches parametrisation model. All fits were performed with DIS data.}
\label{table:blackjax-vs-ultranest}
\end{center}
\end{table}
For these settings, BlackJAX is over four times faster than UltraNest on CPU and 174 times faster on GPUs. Furthermore, the slow GPU-CPU transfers needed by UltraNest make it slower on GPU than it is on CPU.

Note that these comparisons are expected to vary depending on the amount of data included, the parametric form, and the number of free parameters, so is not applicable to other models. Please refer to Ref.~\cite{yallup2025nested} for a more exhaustive quantitative comparison between BlackJAX and other nested samplers.

UltraNest remains available within Colibri, such that the user has the flexibility to choose between them, as it has slightly different features from BlackJAX that might be desirable in some cases, and the possibility to compare results from two different samplers can prove useful.

\section{Conclusion and outlook}
\label{sec:conclusion-outlook}

\noindent Having successfully been used for a variety of studies, such as the exploration of alternative parametrisations for Bayesian PDF determinations~\cite{Candido:2024hjt, Costantini:2025wxp}, or an analysis of the training dynamics of the \texttt{n3fit} neural-network architecture~\cite{chiefa2026}, Colibri is proving to be a useful tool to address the inverse problem of PDF fitting from different angles. 

In these proceedings we have presented two new updates to the code, namely the benchmark and public release of the \texttt{colibri-n3fit} model, based on the NNPDF \texttt{n3fit} architecture, and the implementation of the BlackJAX nested sampler algorithm. These updates contribute to making Colibri a more robust and flexible tool, enabling a wide range of studies.

On one hand, the public availability of the \texttt{colibri-n3fit} model opens the door to quantitative Bayesian studies of the \texttt{n3fit} architecture, and to model comparisons with other implemented models. On the other, access to the BlackJAX nested sampler allows efficient Bayesian inference, bridging the gap with the computational cost of other techniques such as the MC replica or Hessian methods.

A further advantage of adopting BlackJAX is that it is not just a nested sampler, but a Bayesian inference library, meaning that it provides alternative Bayesian inference techniques. Work is ongoing to implement other algorithms. 

Other frontiers, such as the possibility of performing simultaneous fits of Standard Model and Beyond the Standard Model parameters, or the benchmarking and publication of a realistic polynomial model, are being pursued to extend the capabilities of Colibri for Bayesian studies of PDFs.
\section{Acknowledgements}
\label{sec:acknowledgements}

\noindent We are grateful to Luca Mantani and Mark Nestor Costantini for feedback, reviews and input. We would also like to thank Nina Elmer and Maria Ubiali for helpful comments on these proceedings. Likewise, we would like to thank Juan Cruz Martinez for assistance in the Colibri-n3fit implementation. Finally, we extend our gratitude to David Yallup for his time and input in the implementation of BlacJAX in Colibri, as it would not have been possible without him.

\bibliographystyle{utphys}
\bibliography{bib}

\end{document}